\documentclass[twocolumn,twoside,slac_two]{revtex4}
\usepackage{graphicx}
\usepackage{fancyhdr}
\renewcommand{\headrulewidth}{0pt}
\renewcommand{\footrulewidth}{0pt}

\begin{document}

\title{Thermal emission in the prompt phase of gamma-ray bursts}

%

\author{F. Ryde}
%
\affiliation{Stockholm Observatory, AlbaNova, SE-106 91 Stockholm,
Sweden}
\

\begin{abstract}
I discuss the interpretation of the prompt phase in gamma-ray
bursts as being dominated by quasi-thermal emission, rather than
by synchrotron emission. Such an interpretation gives a more
natural explanation of (i) the observed variety of spectral shape
and spectral evolution, (ii) the observed narrowness of the
distribution of peak energies, as well as (iii) the observed
correlations between peak energy and luminosity. However, the
physical setting that could produce such a scenario is not yet
clear.

\end{abstract}

\maketitle

\thispagestyle{fancy}


\section{Introduction}
The prompt emission of gamma-ray bursts (GRBs), radiating mainly
as gamma-rays and X-rays, has defied any simple explanation,
despite the presence of a rich observational material and great
theoretical efforts. This is in contrast to the afterglow
emission, in many cases detected all the way from X-rays to radio
wavelengths, which is successfully described by synchrotron
emission from a forward shock moving at great speed into the
surrounding medium. Synchrotron emission is also a natural
candidate for the prompt emission since it arises naturally in an
ultra-relativistic outflow in which the kinetic energy is
dissipated through, for instance shocks or magnetic reconnections,
and shared between the magnetic field and particles. However,
there are several observational facts that contradict such a
simple picture, most importantly the existence of spectra which
are much too hard, see for instance the spectra from GRB930214 in
Figure 1 and GRB960530 in Figure 2. In many cases such spectra are
fitted well by a thermal emission function \cite{ghirlanda,
kaneko, ryde04}. Furthermore, Ryde \cite{ryde05} showed that
spectra from more typical bursts, that is, bursts having spectra
which are consistent with the synchrotron model, can indeed be
fitted with a hybrid model which is dominated by a thermal
component, but that is overlayed with a non-thermal emission
component as well. In many cases such a model gives a
statistically better fit. Such an example is given in Figure 3.
Even though bursts appear to have a variety, sometimes complex,
spectral evolutions, the behavior of the two separate components
is remarkably similar for all bursts, with the temperature
describing a broken power-law in time. The non-thermal component
is, in most cases, consistent with emission from a population of
fast cooling electrons emitting optically-thin synchrotron
emission or non-thermal Compton radiation, giving a power-law
slope of the photon spectrum of $s = -1.5$. However, in the case
of GRB960530, shown in Figure 2, $s$ is closer to $-2/3$, which is
expected for slow cooling \cite{RB}.

It is very important to note that it is the {\it instantaneous}
spectra of GRBs that most closely should reveal the radiation
mechanism. This is because of the strong spectral evolution that
normally occurs during a burst and that will make the spectral
shape of the {\it time-integrated} spectrum differ from that given
by the emission process. This is in particular the case for
complex bursts with several emission peaks. The time-integrated
spectrum can easily be found from the instantaneous spectra and
the spectral evolution, which was shown analytically by Ryde \&
Svensson \cite{RS99}; the spectra always become softer.

\begin{figure*}[]
\includegraphics[width=135mm]{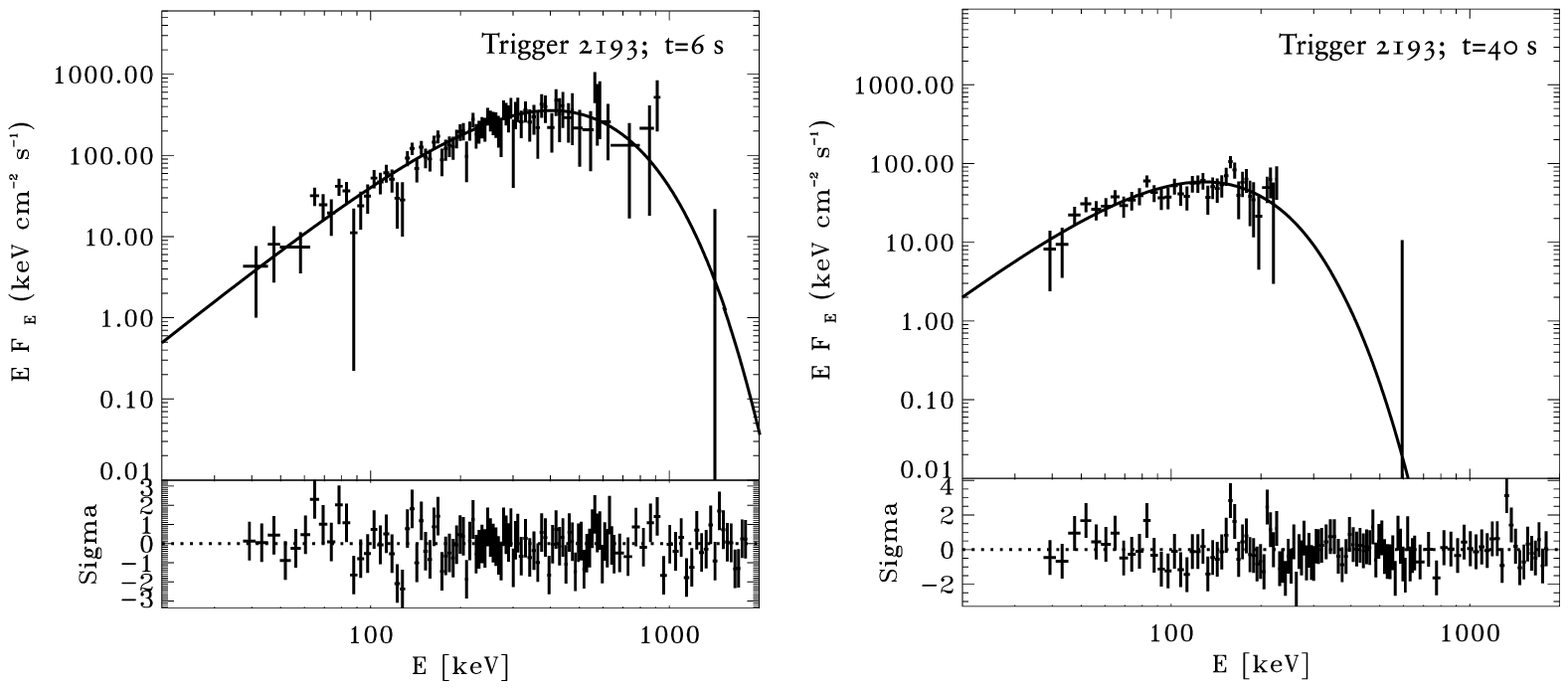}
\caption{Two time-resolved spectra from GRB930214 (BATSE trigger
2193) from 6 and 40 seconds after the trigger. Note that the
spectra are fitted well with a Planck function, both in the
Rayleigh-Jeans portion of the spectrum with $\alpha=+1$ and in the
Wien portion with a fast voidance of flux. The temperature has
changed between the measurements.} \label{fig:f1}
\end{figure*}

\section{Synchrotron Model}

It was early recognized that the spectra of gamma-ray bursts
(GRBs) have a non-thermal character, with emission over a broad
energy range (e.g. \cite{FM}). This typically indicates emission
from an optically-thin source and an initial proposal for GRBs was
therefore an optically-thin synchrotron model from
shock-accelerated, relativistic electrons (e.g. \cite{katz,
Tavani}). The number density of the radiating electrons is assumed
to be typically a power law as a function of the electron Lorentz
factor $\gamma_{\rm e}$ above a  minimum value, $\gamma_{\rm
min}$, with index $-p$. Such a distribution gives rise to a
power-law photon spectrum with photon index $\alpha = -2/3$ below
a break energy $E_{\rm p} \propto \gamma _{\rm min} ^2$ and a
high-energy power-law with index $\beta = -(p+1)/2$. However, as
mentioned above, this model has difficulties in explaining the
observed spectra of GRBs, which show a great variation in $\alpha$
and $\beta$ (see \cite{preece00}). In particular, a substantial
fraction of them have $\alpha > -2/3$, which is not possible in
the model in its simplest form, since $\alpha = -2/3$ is the
power-law slope of the fundamental synchrotron function for
electrons with an isotropic distribution of pitch angles
\cite{pac}. The problem becomes even more severe for the case when
the cooling time of the electrons is shorter than the typical
dynamic timescale. In the typical setting of GRBs having a
relativistic outflow with a bulk Lorentz factor $\Gamma \sim 100$,
the time scales for synchrotron and inverse Compton losses are
$\sim 10^{-6}$ s \cite{ghis00}, which is much shorter than both
the dynamic time scale $R/2\Gamma^2 c \sim 1 \,\, {\rm s} \,\,(R /
10^{15}\, \,{\rm cm})$, and the integration time-scale of the
recorded data, typically 64 ms to 1 s. In such a case the
low-energy power-law should be even softer, with $\alpha = -1.5$
\cite{bussard, g00}, now contradicting a majority of the observed
spectra. Furthermore, the observed distribution of $\alpha$ from
{\it instantaneous} spectra is smooth (see \cite{preece00}) and
does not show any indication of preferred values, such as $-2/3$
or $-3/2$. Other variations of the synchrotron or/and inverse
Compton model have been suggested (see e.g.
\cite{BaringB,LP00,SP04}) to account for these hard spectra.

The peak energy from the above distribution of electrons is given
by $E_{\rm pk} = \gamma_{\rm m}^2 B_\perp \Gamma$. In the external
shock model $\gamma_{\rm m}$ and  $B_\perp$ are proportional to
the bulk Lorentz factor, which makes $E_{\rm pk}\propto \Gamma^4$,
which is a very strong dependence, which poses a problem  in
explaining the relative narrowness of the distribution of peak
energies \cite{preece00}, even including the X-ray flashes. For
the internal shock model the $\gamma_{\rm m} \propto \Gamma_{\rm
rel}$, which is the relative Lorentz factor between the two shells
that collide and
\begin{equation}
E_{\rm pk} \propto B_{\perp} \Gamma. \label{eq:1}
\end{equation}
 However, the sharing of the energy between the  kinetic energy of
the electrons and the magnetic fields should lead to a larger
dispersion.

A third complication arises in explaining the observed correlation
between peak energy and the luminosity which was discussed by
Lloyd-Ronning et al. \cite{LPM} and Amati et al. \cite{amati} (see
also \cite{gamati}); the peak energy is correlated with the
isotropically equivalent energy given by
\begin{equation}
E_{\rm pk} \sim \left( \frac{E_{\rm iso}}{1.2\times 10^{53} {\rm
erg}} \right)^{0.40\pm0.05} \label{eq:2}
\end{equation}
where $E_{\rm iso}$ is
\begin{equation}
E_\gamma = (1-\cos \theta) E_{\rm iso}
\end{equation}
where $E_\gamma$ is the actual gamma-ray energy emitted and
$\theta$ is the jet opening half-angle of the collimated outflow.

\begin{figure*}[t]
\includegraphics[width=135mm]{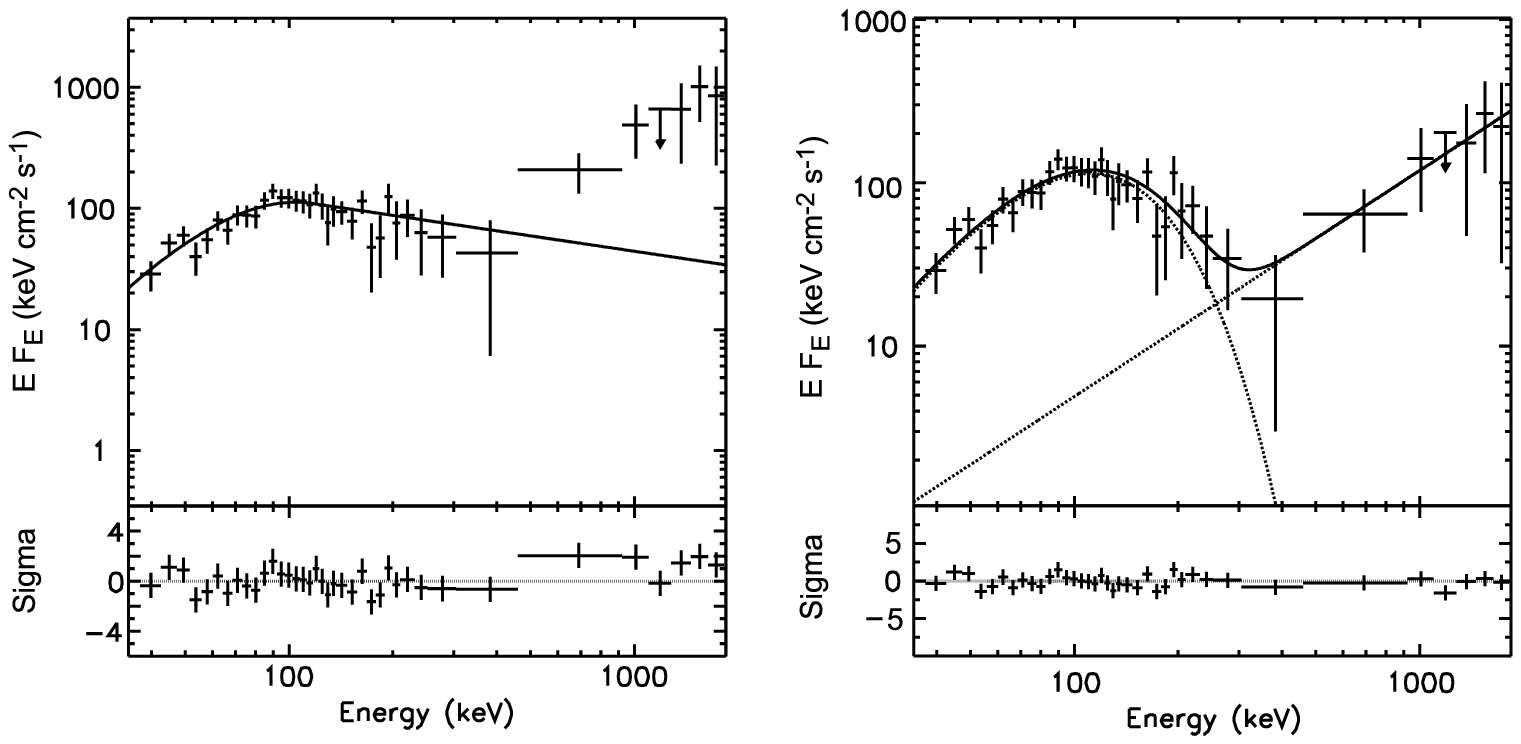}
\caption{\small{The same spectrum from GRB960530 (\#5478; 6 s
after the trigger) fitted with (left panel) the Band et al.
\cite{band93} model with $\alpha = 1.7 \pm 1.5$ and $\beta = -2.4
\pm 0.3$ and (right panel) the hybrid model  \cite{ryde04}, with a
power-law slope of $s = -0.62 \pm 0.27$. Note the obliging of the
data points \cite{fen, BS}.}}
 \label{fig:f1}
\end{figure*}

Equation (\ref{eq:1}) shows that in the internal shock model
$E_{\rm pk}$ is proportional to the Lorentz boosted magnetic field
strength. The total energy density
\begin{equation}
U = \frac{(B\Gamma)^2}{8 \pi} \propto \frac{L}{R^2}
\end{equation}
The typical radius for the internal shocks to occur is $R_{\rm sh}
\sim c t_{\rm v} \Gamma^2$, where $t_{\rm v}$ is the typical
variability time scale, and thus
\begin{equation}
E_{\rm pk} \propto \Gamma^{-2} L^{1/2} t_{\rm v}^{-1}.
\end{equation}
To get a relation similar to that in equation (\ref{eq:2}) both
$\Gamma$ and $t_{\rm v}$ have to be quite similar for all bursts,
which is difficult to imagine. Even though there is no direct way
of determining the bulk Lorentz factor, various physical models
give suggestions on plausible relations between the luminosity and
the bulk Lorentz factor. For instance, \cite{KRM} argued for $L
\propto \Gamma^2$. Such a relation would thus give $E_{\rm pk}
\propto L^{-1/2}$, that is, an anti-correlation, in contradiction
to the observed behavior (see also \cite{RRL,ZM}). Additional
assumptions are needed to explain the positive correlation.
Invoking Poynting flux and/or pair dominated models the
correlation also becomes positive \cite{ZM}.

\section{Quasi-thermal models}

\begin{figure*}[t]
\includegraphics[width=135mm]{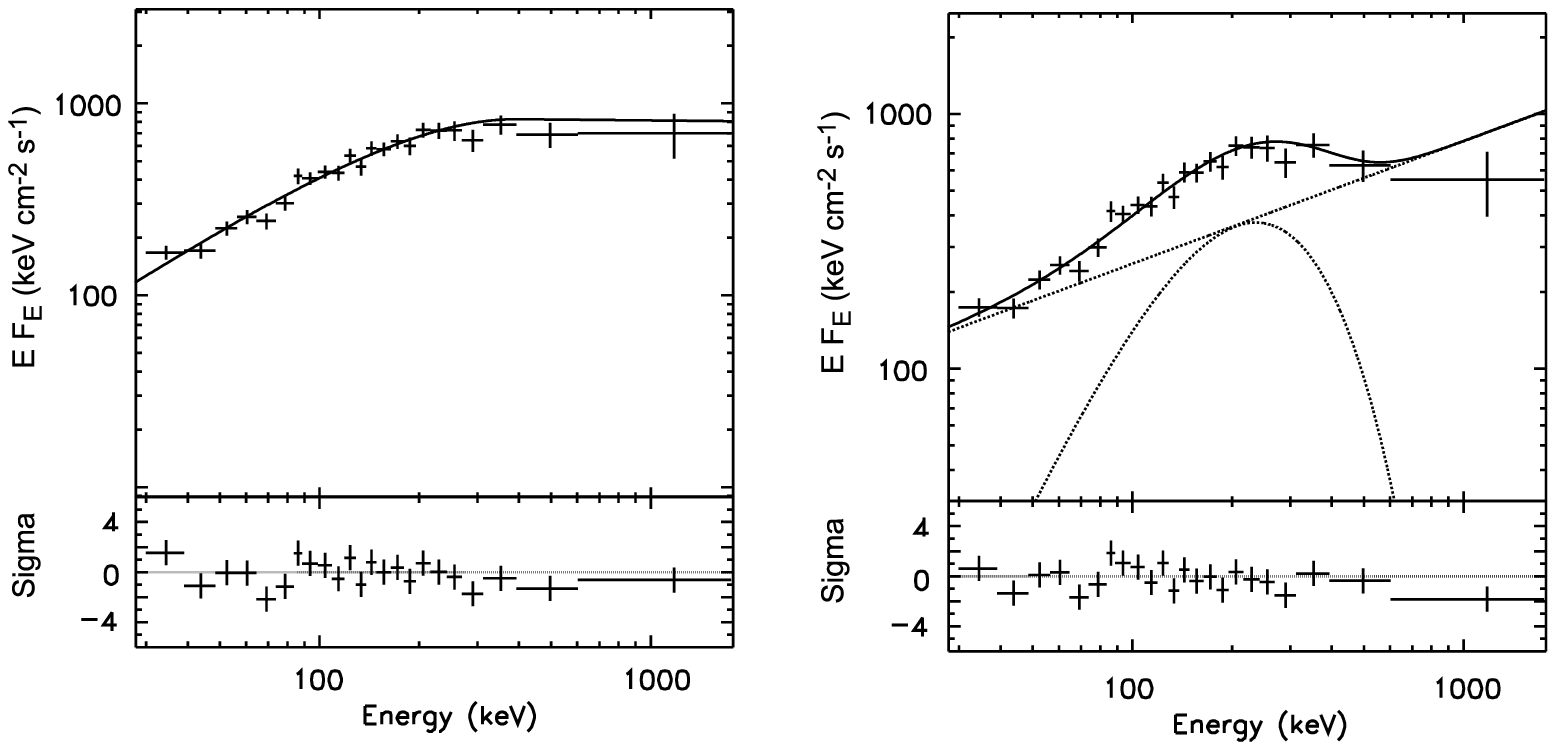}
\caption{\small{The same spectrum from GRB911031 (\#973; 2.5 s
after the trigger) fitted with (left panel) the Band et al.
\cite{band93} model with $\alpha = -0.86 \pm 0.12 $ and $\beta =
-2.4 \pm 0.3$ and (right panel) the hybrid model \cite{ryde04},
with a power-law slope of $s = - 1.52 \pm 0.04$.}}
 \label{fig:f1}
\end{figure*}

If the prompt phase is indeed dominated by a thermal component
these three issues become natural consequences. First, as shown by
Ryde (2004, 2005) the relative strength and the slope of the
non-thermal component will determine the value of the low-energy
power-law index, $\alpha$, that would be found if the Band et al.
\cite{band93} function were to be used. If the thermal component
is strong and/or the non-thermal component is hard, the resulting
spectrum will have a hard $\alpha$ (see Fig.2). While if the
non-thermal component becomes relatively stronger and/or softer
the measured $\alpha$-value would be softer (see Fig. 3). The
observed distribution of $\alpha$-values is therefore consistent
with this picture and in particular the spectra beyond the
"line-of-death" \cite{lod} are not conspicuous. A strong spectral
evolution, for instance, in a large change in the measured value
of $\alpha$, is also easily explained.

Second, the peak of the spectrum is now determined of $kT$ and is
less sensitive to the bulk Lorentz factor. In fact, if the
photosphere occurs during the acceleration phase it is practically
independent of $\Gamma$.  Rees \& M\'esz\'aros \cite{RM05} suggest
a model where the photospheric emission can become enhanced by
dissipative effects below the photosphere (magnetic reconnections,
shocks) and subsequent Comptonization, see also \cite{PW}. Typical
values for the peak energy would be hundreds of keV.

Third, the correlation in equation (\ref{eq:2}) has a natural
explanation since for a thermal emitter the luminosity and the
temperature are correlated. For instance, equating $E_{\rm pk}$
with the energy density, one gets $ E_{\rm pk} \propto \Gamma kT
\propto \Gamma U^{1/4} \propto \Gamma (L/\Gamma R^2)^{1/4}.$ Using
that the pair photosphere occurs at $R_{\rm ph} \propto L
\Gamma^{-3}$ (see Rees \& M\'esz\'aros  \cite{RM05}) then $E_{\rm
pk} \propto L^{\sim 0.8}$.
In the last step, $L \propto \Gamma ^2$ was again assumed.
Similarly if the photosphere is emitted during the acceleration
phase its temperature will be constant in the observer frame since
the comoving cooling by adiabatic expansion is compensated for by
the increase in $\Gamma$.
\begin{equation}
\Gamma kT_0 \propto \left( \frac{L}{R_0^2}\right)^{1/4} \propto
L^{1/4} R_0^{-1/2}.
\end{equation}
Here, $R_0$ is the radius at which the linear acceleration starts.
Assuming, for instance, that $R_0 \propto L^{-1}$ then again
$E_{\rm pk} \propto L^{0.75}$ (see further Rees \& M\'esz\'aros
\cite{RM05}). Finally, for extremely photon starved plasmas
$E_{\rm pk} \propto L/N_\gamma\propto L$.

\section{Conclusions}

The radiative efficiency of the thermal emission can in plausible
scenarios be radically increased by, for instance, dissipation
processes below the photosphere \cite{RM05, PW}. These processes
would naturally produce large amount of electron-positron pairs
with modest Lorentz factors, which would Compton up-scatter the
thermal radiation. The observed peak would then be this
Comptonized peak. The non-thermal emission seen in the spectra,
could be due to synchrotron emission or inverse Compton emission
from dissipation regions outside the photosphere. Ryde
\cite{ryde05} showed that the energy flux in the thermal and the
non-thermal components are correlated which might indicate the
latter.

In summary, thermal emission could indeed dominate over
non-thermal emission in standard settings of a GRB jet. In such
scenarios a correlation between the peak energy and the luminosity
naturally arises, the details somewhat depending on the
dissipation processes. In addition, the dispersion in $E_{\rm pk}$
would be smaller and the observed spectral shapes and spectral
evolution get natural explanations.


\begin{thebibliography}{9}

\bibitem{ghirlanda}Ghirlanda, G.,  Celotti, A., \& Ghisellini, G.
        2003, A\&A, 406, 879

\bibitem{kaneko}Kaneko, Y., Preece, R. D., \& Briggs, M. S. 2003, AAS, 203, 8004

\bibitem{ryde04} Ryde, F. 2004, ApJ, 614, 827

\bibitem{ryde05}Ryde, F. 2005, ApJ Letters, in press
(astro-ph/0504450)

\bibitem{RB}Ryde, F., \& Battelino, M. 2005, submitted

\bibitem{RS99}Ryde, F. \& Svensson, R. 1999, ApJ, 512, 693

\bibitem{FM}Fishman, G.J., \& Meegan, C.A. 1995, ARAA, 33, 415

\bibitem{katz}Katz, J. I. 1994, ApJ, 432, L107

\bibitem{Tavani}Tavani, M. 1996, ApJ, 466, 768

\bibitem{preece00} Preece, R. D., Briggs, M. S.,
Mallozzi, R. S., Pendleton, G. N., Paciesas, W. S., \& Band, D. L.
2000, ApJSS, 126, 19

\bibitem{pac}Pacholczyk, A. G. 1970, {\it Radio
Astrophysics} (San Francisco: W. H. Freeman and Co.)

\bibitem{ghis00} Ghisellini, G., Lazzati, D., Celotti, A., \& Rees, M. J.
2000, MNRAS, 316, L45

\bibitem{bussard}Bussard, R.W., ApJ, 284, 357

\bibitem{g00}Ghisellini, G., Celotti, A., Lazzati, D. 2000, MNRAS 313, L1

\bibitem{BaringB}Baring, M.G., \& Braby, M.L. 2004, ApJ, 613,
460

\bibitem{LP00}Lloyd-Ronning, N. \& Petrosian, V. 2000, ApJ, 543, 722

\bibitem{SP04} Stern, B. \& Poutanen, J. 2004, MNRAS, in press

\bibitem{LPM}Lloyd-Ronning, N. M., Petrosian, V., \& Mallozzi, R. S. 2000, ApJ, 534, 227

\bibitem{amati} Amati et al. 2002, A\&A, 390, 81

\bibitem{gamati}Ghirlanda, G., Ghisellini, G., \& Lazzati, D. 2004, ApJ, 616, 331

\bibitem{fen}Fenimore, E. E., Klebesadel, R. W., \& Laros, J. G. 1983, in Gamma-Ray
Astronomy in Perspective of Future Space Experiments (New York:
Pergamon), 201

\bibitem{BS}Bromm, V. \& Schaefer, B. 1999, ApJ, 520, 661

\bibitem{KRM}Kobayashi, S., Ryde, F., \& MacFadyen, A. 2002, ApJ, 577, 302

\bibitem{RRL}Ramirez-Ruiz, E., \& Lloyd-Ronning, N.M. 2002, NewA,
7, 197

\bibitem{ZM}Zhang, B. \&  M\'esz\'aros, P. 2002, ApJ, 581, 1236

\bibitem{band93} Band, D., et~al.\ 1993, ApJ,  413, 281

\bibitem{lod}Preece, R.D., Briggs, M.S., Mallozzi, R.S., Pendleton,
G.N., Paciesas, W.S., \& Band, D.L. 1998, ApJ, 506, L23

\bibitem{RM05}Rees, M. J., \& M\'esz\'aros, P. 2005, submitted

\bibitem{PW}Pe'er, A., \& Waxman, E. 2004, ApJ, 613, 448

\end{thebibliography}
\end{document}